\documentclass[prd,preprintnumbers,amsmath,amssymb,superscriptaddress,nofootinbib]{revtex4}
\usepackage{graphicx}
\usepackage{dcolumn}
\usepackage{bm}
\usepackage{graphicx,amsmath,amssymb,color}
\usepackage{relsize}
\usepackage{slashed}
\usepackage{graphicx}
\usepackage{bm}
\usepackage{mathtools}
\usepackage{mathrsfs}
\usepackage{soul}
\usepackage{verbatim} 
\usepackage{tikz}
\usepackage{array}
\usepackage{float}
\usepackage{hyperref}
\usepackage[caption=false]{subfig}

\allowdisplaybreaks
\def\be{\begin{eqnarray}}
\def\ee{\end{eqnarray}}
\def\bea{\begin{eqnarray}}
\def\eea{\end{eqnarray}}
\def\bes{{\bf K}}

\begin{document}

\title{Angular Momentum Distribution in the Transverse Plane}
\author{Lekha Adhikari}\email{adhikari@iastate.edu}
\affiliation{Department of Physics and Astronomy, Iowa State University,
Ames, IA 50011, U.S.A.}
\author{Matthias Burkardt}\email{burkardt@nmsu.edu}
\affiliation{Department of Physics, New Mexico State University,
Las Cruces, NM 88003, U.S.A.}

\date{\today}

\begin{abstract}
Several possibilities to relate the $t$-dependence of Generalized Parton Distributions (GPDs) to the distribution of angular momentum in the transverse plane are discussed. Using a simple spectator model we demonstrate that
non of them correctly describes the orbital angular momentum distribution that for a longitudinally polarized nucleon obtained directly from light-front wavefunctions.
\end{abstract}
\maketitle
\section{Introduction}
Since the famous EMC experiment \cite{Ashman:1987hv} there has been great interest to understanding the contributions from orbital angular momentum and from gluon spin to the nucleon spin.
Meanwhile, Generalized Parton Distributions (GPDs) have been introduced as a novel tool to describe the internal structure of hadron. Over more than a decade, there has been a strong interest in GPDs as many observables can be linked to them. Specifically, GPDs have been used extensively after they were first identified with the total angular momentum of the quarks or gluons within a nucleon. The total angular momentum carried by the quarks is calculated using the $2^{\text{nd}}$ moment of GPDs as  \cite{Ji:PRL}
\begin{equation}
J^z = \frac{1}{2}\int dx\,x\left[ H(x, 0, 0)
+E(x, 0, 0)\right]. \label{eq:Jirelation}
\end{equation}
However, this famous Ji relation or sum rule, yields the $z$ component of the total angular momentum of the quarks
in a nucleon that is polarized in the $+z$ direction only when GPDs are extrapolated to $t=0$ in the general expression
\begin{equation}
J(t) \equiv \frac{1}{2}\int dx\,x\left[ H(x,\xi,t)
+E(x,\xi,t)\right], \label{eq:Jqt}
\end{equation}
where $t=\Delta^2$, $x$ is the light-cone momentum fraction carried by the quark
(averaged between initial and final state), $\xi \equiv \frac{p^+-p'^+}{p^++p'^+} = -\frac{\Delta^+}{p^++p'^+}$ is the longitudinal momentum transfer, $p$ and $p'$ are initial and final state momenta of the nucleon, respectively, and $\Delta \equiv p'-p$ is the  momentum transfer.

GPDs have also been used to visualize nucleons in three-dimensions after doing the suitable Fourier transform (FT) of these GPDs \cite{Burkardt:2002hr, mb:GPD,Diehl:2002he}.  These images in a space where one dimension describes the light-cone momentum fraction ($x$) and the other two dimensions describe the transverse position (${\vec b}_\perp$) of the parton (relative to the transverse center of momentum). This distribution of partons in the transverse plane has a probabilistic
interpretation, in the same sense and with the same limitations as the usual parton distributions. For an unpolarized nucleon,  the $2$-dimensional FT of GPD $H(x,0,t)$ in the transverse plane reads \cite{mb:GPD}
\be  q(x, {\vec b}_\perp) =
\int \frac{d^2{\vec \Delta}_\perp}{(2\pi)^2}
e^{-i {\vec \Delta}_\perp \cdot {\vec b}_\perp }
H(x,0,-\vec{\Delta}_\perp^2),\label{eq:gpds_space} \ee
where the impact parameter ${\vec b}_\perp$, which is the Fourier-conjugate to ${\vec \Delta}_\perp$, is defined in the 2-dimensional transverse plane perpendicular to the light-cone direction.  Here, $ |{\vec b}_\perp|\equiv b_\perp$ is introduced as the displacement of the active quark $(q)$ from the transverse center of momentum  of the entire nucleon. The transverse center of momentum ${\vec R}_\perp$ is defined as
the weighted average of the transverse positions of all partons, where the weight factor is the respective momentum fraction
 \cite{Soper:1976jc}
\be {\vec R}_\perp  = \sum_{i\in q,g}x_i{\vec r}_{\perp i}
= x{\vec r}_{\perp}+(1-x){\vec R}_{\perp s}
,\ee
where $x$ and ${\vec r}_\perp\equiv {\vec r}_{\perp 1}$ are the momentum fraction and transverse position of the active quark and $1-x$ and ${\vec R}_{\perp s}$ are that
of the spectator(s).
Therefore, for a quark, one can write
\be {\vec b}_\perp = {\vec r}_{\perp } -{\vec R}_\perp = (1-x)({\vec r}_\perp-{\vec R}_{\perp s}). \label{eq:bperp}\ee
Further details of $q(x, {\vec b}_\perp)$ can be found in Refs.~\cite{Burkardt:2002hr, mb:GPD,Diehl:2002he, Diehl:2003ny}.

In the context of $t\not= 0$, there have been many discussions on the connection between the current theoretical GPD framework and Deeply Virtual Compton Scattering (DVCS) experiments. One can find the details in Ref. \cite{Bakker:2011zzb}. It is therefore  useful to study  the $t$-dependence of GPDs with the total angular momentum in the coordinate space to check if the partonic interpretation still holds in that space. Furthermore, the work presented in Ref.~\cite{Polyakov:2002yz} (see also Ref.~\cite{Goeke:2007fp}) suggests that the 3-dimensional FT of Eq.~(\ref{eq:Jqt}) can be used to calculate the distribution of angular momentum in coordinate space. The `Chiral Quark Soliton Model' that was used in Ref. \cite{Goeke:2007fp} had an infinite target mass, and therefore there was no issue of relativistic corrections. The relativistic corrections would potentially be an issue upon taking the 3-dimensional  FT of the generalized form factors [available in Eq.~(\ref{eq:Jqt})] for
finite nucleon mass. The interpretation of the 3-dimensional FT of form factors (FFs) as a distribution in 3-dimensional space becomes ambiguous and suffers from relativistic corrections at length scales equal to or smaller than the Compton wavelength of the target \cite{mb:GPD, Miller:2007uy}. However, a 2-dimensional FT of FFs does not suffer from such relativistic corrections. As a corollary, one finds that the distribution of charge in the transverse plane is given by the $2$-dimensional FT of the Dirac form factor \cite{Miller:2007uy}. Bcause of these results, it has been suggested (see for example Ref. \cite{Burkert:2012rh}) to perform a 2-dimensional FT of FFs to study the angular momentum distribution in the transverse plane.
Furthermore, there are potential issues due to total derivative terms linking canonical and symmetric energy momentum tensors \cite{Leader:2013jra}.
We are therefore motivated by these investigations and suggestions as mentioned above to investigate the $t$-dependence of GPDs with the total angular momentum  distributions of the quark in the transverse plane for a longitudinally polarized nucleon.
Before proceeding, we should note that while $L_z$ does not commute with ${\vec b}_\perp$, it does
commute with $|{\vec b}_\perp|$ and it is thus meaningful to discuss $L_z(|{\vec b}_\perp|)$.

In the following sections, we prescribe four different techniques to study the angular momentum distributions in the context of the Scalar Diquark Model (SDQM). First, we define a 2-dimensional FT of Eq.~(\ref{eq:Jqt}) as $\widetilde{J}({\vec {b}_\perp})$ \cite{Adhikari:2013ima}. This technique will be referred to as naive technique in this paper. In the second technique, using $\widetilde{J}({\vec b}_\perp)$, we derive the 2-dimensional FT of the result that was originally suggested for the purpose of  the 3-dimensional FT in Refs.~\cite{Polyakov:2002yz,Goeke:2007fp}. This technique will be referred to as Polyakov-Goeke (PG) technique. In the third technique, we present an independent derivation of the distribution of angular momentum in the transverse plane in the Infinite Momentum Frame (IMF). Finally, we will compare these three different distributions with the one  calculated  directly from light-front wavefunctions (LFWFs) using widely-recognized Jaffe-Manohar (JM) definition for orbital angular momentum (OAM). In momentum space, this definition preserves the partonic interpretation.

\section{Distribution of Angular momentum in the transverse plane}
\subsubsection{\textsc{\bf Naive Technique:}}
 We take the 2-dimensional FT of Eq.~(\ref{eq:Jqt}) to calculate the distribution of total angular momentum (TAM) in the transverse plane in the Drell-Yan frame as
\be\widetilde{J}(\vec b_\perp) = \int \! \frac{d^2{\vec \Delta}_\perp}{(2\pi)^2} e^{-i {\vec \Delta}_\perp \cdot \vec{ b}_\perp  } J(t=-\vec \Delta^2_\perp) \label{eq:TAMnaive}
\ee
where the TAM distribution depends only on the distance $b_\perp \equiv |{\vec b}_\perp|$ from the origin, which is the transverse center of the entire nucleon.
It is evident that
\be
\int d^2{\vec b}_\perp  \widetilde{J}({\vec b}_\perp)= J(t=0) \equiv J^z,
\ee
but that does not automatically imply that $\widetilde{J}(\vec b_\perp)$ represents the distribution of TAM
in the transverse plane.
\subsubsection{\textsc {\bf Polyakov-Goeke (PG) Technique: Two Dimensional Reduction of Polyakov-Goeke Prescription:}}
The relation between the symmetric energy-momentum tensor (EMT) and the total angular momentum distribution, which is defined and available in Ref.~\cite{Polyakov:2002yz}, reads  \cite{Pagels:1966zza,Ji:PRL, Ji:1996nm}
\be J(t) + \frac{2t}{3}\frac{d}{dt}J(t) = \int\!\! d^3{\vec b}\, e^{i\vec{b}\cdot \vec{\Delta}}\,\epsilon^{ijk} s_i b_j T_{0k}(\vec {b},\vec{s}), \label{eq:Goeke9}
\ee
where $T^{\mu \nu}$ is the EMT, $T_{0k}(\vec{b},\vec{s})$ represents the momentum distribution of the quarks within the nucleon and thus the integrand on the right hand side is interpreted as the angular momentum density, ${\vec b}$ is 3-dimensional coordinate space, and $\vec s$ is the nucleon spin. In our case, the nucleon is polarized in the $+z$ direction.

Refs.~\cite{Polyakov:2002yz,Goeke:2007fp} suggest that the  3-dimensional FT of Eq.~(\ref{eq:Goeke9}) yields the distribution of angular momentum in 3-dimensional coordinate space.  As mentioned before, there is an issue of relativistic corrections for the  3-dimensional FT. However, there are no such corrections to the  2-dimensional FT of FFs in the infinite momentum frame. We thus define the 2-dimensional FT in the coordinate space (${\vec b}_\perp$) as
\be \rho_J^{\text{PG}}({\vec b}_\perp) \!\equiv \!\int \!db_z \,\epsilon^{ijk}\,s_i\,b_j\, T_{0k}(\vec {b},\vec{s})
.\ee
 With the help of Eq.~(\ref{eq:TAMnaive}), the 2-dimensional FT of Eq.~(\ref{eq:Goeke9})  yields
\be    \rho_J^{\text{PG}}(b_\perp) = \frac{1}{3}\widetilde{J}( b_\perp) -\frac{1}{3} b_\perp \frac{d}{db_\perp} \widetilde{J}(b_\perp).  \label{eq:TAMGoeke} \ee
In the following, the total angular momentum (TAM) distribution defined by Eq.~(\ref{eq:TAMGoeke}) will be referred to as Polyakov-Goeke (PG) technique because it was derived as the  2-dimensional reduction of the result available in Ref. \cite{Polyakov:2002yz}.
\subsubsection{\textsc {\bf Infinite Momentum Frame (IMF) Technique:}}
As an alternative, one can  derive the distribution of TAM directly in the Infinite Momentum Frame (IMF). For this purpose, we define the EMT $T^{\mu \nu}$ in terms of several form factors. Note that it is the same $T^{\mu \nu}$ that was  used to derive Eq.~(\ref{eq:Goeke9}). The form factors of the symmetric EMT (valid for quarks/gluon) read \cite{Ji:PRL, Goeke:2007fp}
\be \langle p^\prime | T^{\mu \nu} |p\rangle \!\! = \bar{u}(p^\prime)\biggr[ M_2(t) \frac{P^\mu P^\nu}{M_N}
+ J(t) \frac{i(P^\mu \sigma^{\nu \rho}+P^\nu \sigma^{\mu \rho})}{2M_N}
\Delta_\rho +   d_1(t) \frac{\Delta^\mu \Delta^\nu -g^{\mu \nu} \Delta^2}{5M_N} \pm \bar{c}(t)g^{\mu \nu} \biggr]u(p),\label{eq:Tmunu}
\ee
where 
$u(p)$ is the nucleon spinor, $M_N$ is the mass of the nucleon, and $2P=p+p^\prime$.

The $z$-component of the total angular momentum of the quarks  in terms of angular momentum density $M^{\alpha \mu \nu}$ is defined as \cite{Jaffe:1989jz}
\be  J^z =  \epsilon^{zxy} \int\! d^2{\vec b}_\perp \, M^{+xy}, \!\!\quad M^{+xy}\! =\! T^{+y}b_x -T^{+x}b_y. \ee

To define TAM distribution in the transverse plane, one needs to localize the nucleon in the transverse direction. For this purpose, states $|p^+, {\vec R}_\perp, s_z \rangle $, which are the eigenstates of the transverse center of momentum, are introduced in terms of the light-cone helicity eigenstates as \be  |p^+, {\vec R}_\perp = {\vec 0}_\perp, s_z \rangle = \frac{\mathcal N}{(2\pi)^2} \int  d^2{\vec p}_\perp  |p^+,{\vec p}_\perp,s_z \rangle,\label{eq:wavepacket} \ee
where ${\mathcal N}$ is a normalization constant satisfying $|{\mathcal N}|^2 \int \frac{d^2{\vec P}_\perp}{(2 \pi)^2} =1$ \cite{Diehl:2003ny,Burkardt:2002hr,Soper:1976jc, mb:GPD, Diehl:2002he}.

For these transversely localized states
the TAM distribution in impact parameter space (in the IMF) can be defined as
\be \rho^{\text{IMF}}_J({\vec b}_\perp)\equiv  \frac{1}{p^+} \langle p^+, {\vec R}_\perp = {\vec 0}_\perp, s_z|\left[T^{+y}({\vec b}_\perp)b_x \right. - \left.T^{+x}({\vec b}_\perp)b_y\right]|p^+, {\vec R}_\perp = {\vec 0}_\perp,s_z \rangle.  \label{eq:rhodef}
\ee
In order to evaluate the matrix elements we note the phase factor
\be \langle p^+,{\vec p}^{\prime}_\perp,s_z|T^{+j}({\vec b}_\perp)|p^+,{\vec p}_\perp,s_z \rangle =
 \langle p^+, p^{\prime}_\perp,s_z|T^{+j}(0_\perp)|p^+, p_\perp,s_z \rangle  e^{-i\vec b_\perp\cdot \vec\Delta_\perp}, \quad{j=x,y},\label{eq:rhophase}
\ee
where the matrix elements of the EMT are evaluated using Eq.~(\ref{eq:Tmunu}). We note that only the
term involving $p^+\sigma^{xy}\Delta^y$ in
second term of Eq.~(\ref{eq:Tmunu}) survives taking the matrix elements between helicity eigenstates
that have the same light-cone helicity $[ \langle p^+,p^{\prime}_\perp,s_z|\sigma^{+j}|p^+,p_\perp,s_z \rangle=0]$ and symmetric integration around the $z$ axis.

Eq.~(\ref{eq:rhodef}) after simplification thus yields
\be  \rho^{\text{IMF}}_J(b_\perp)=\mp \frac{1}{2} \left [ b_\perp \frac{d}{db_\perp} \widetilde{J}(b_\perp) \right],  \label{eq:TAMIMF} \ee
where the TAM distribution depends only on the distance $b_\perp \equiv |{\vec b}_\perp |$.

Here, it is noted that Eq.~(\ref{eq:Goeke9}), which was derived in Ref.~\cite{Polyakov:2002yz}, has been used in Ref.~\cite{Goeke:2007fp} to study the angular momentum density in the infinite target mass frame. Eq.~(\ref{eq:Goeke9}) was later modified to its relativistic version in Ref.~\cite{Leader:2013jra}. One can derive Eq.~(\ref{eq:TAMIMF}) directly from the relativistic version of Eq.~(\ref{eq:Goeke9}) available in Ref.~\cite{Leader:2013jra}.

\section{model calculations and results}
We use the SDQM to calculate the proposed OAM densities to test if any of the densities agrees with the distribution of OAM in the transverse plane obtained using light-front wavefunctions which has
a partonic interpretation.
The SDQM is not a good approximation for Quantum Chromodynamics (QCD). However, it is perfect to illustrate a point-of-principle:
none of the above proposed distributions agree with the partonic calculation to be derived below. What is particularly useful about the SDQM in this context is that maintaining
Lorentz invariance (which is important here) is straightforward and there are no added complications due to the absence of gauge fields. Of course QCD is a gauge theory, but if a certain interpretation fails already in a non-gauge theory - as we will demonstrate - it is very unlikely to hold in a gauge theory.

The relevant GPDs contained in Eq.~(\ref{eq:Jqt}) are calculated using the following LFWFs  for the SDQM \cite{Brodsky:2000xy, Brodsky:1980zm,Brodsky:1997de}:
\be
\begin{split}
\psi_{+\frac{1}{2}}^\uparrow \left(x,{\vec k}_\perp\right)
&= \left(M+\frac{m}{x}\right) \phi (x,{\vec k}_\perp^2),\quad
\psi_{-\frac{1}{2}}^\uparrow (x,{\vec k}_\perp)&=-\frac{k^1+ik^2}{x} \phi (x,{\vec k}_\perp^2),\\
 \psi^\downarrow_{+\frac{1}{2}}(x,{\vec{k}}_\perp)&=\frac{k^1-ik^2}{x}\phi (x,{\vec k}_\perp^2),\quad \psi^\downarrow_{-\frac{1}{2}}(x,{\vec{k}}_\perp)&=(M+\frac{m}{x})\phi (x,{\vec k}_\perp^2),
\label{eq:SDQM} \end{split}
\ee
where \be \begin{split} \phi ({\vec k}_\perp^2) &\equiv  \phi (x,{\vec k}_\perp^2) = \frac{g/\sqrt{1-x}}{M^2-\frac{{\vec k}_\perp^2+m^2}{x}
-\frac{{\vec k}_\perp^2+\lambda^2}{1-x}} = \frac{-gx \sqrt{(1-x)}}{{\vec k}^2_\perp+u(\lambda^2)} \\  & \text{and}\,\,u(\lambda^2)= m^2(1-x) - M^2 x (1-x)+x{\lambda}^2. \label{eq:phiu} \end{split} \ee
Here, $g$ is the Yukawa coupling and $M$, $m$, and $\lambda$ are the masses
of the `nucleon', `quark', `diquark', respectively. Furthermore,
 ${\vec k}_\perp\equiv
{\vec k}_{\perp q}- {\vec k}_{\perp \text {scalar}}$ represents the relative transverse momentum between the quark and the diquark (scalar). The upper wave function index $\uparrow$ refers to the helicity of the `nucleon' and the
lower index to that of the `quark'.

GPDs using overlap integrals of LFWFs in the Drell-Yan frame read \cite{Brodsky:2000xy}
\vspace{0.6cm}
\be H(x,0,-{\vec \Delta}^2_\perp)&=&\int\!\frac{d^2{\vec{k}}_\perp}{16\pi^3}\biggr[\psi^{\uparrow\,*}_{+\frac{1}{2}}(x,{\vec{k}}^\prime_\perp)\psi^\uparrow_{+\frac{1}{2}}(x,{\vec{k}}_\perp)
 +\psi^{\uparrow\,*}_{-\frac{1}{2}}(x,{\vec{k}}^\prime_\perp)\psi^\uparrow_{-\frac{1}{2}}(x,{\vec{k}}_\perp)\biggr] \label{eq:HGPDs} \\
&=&\frac{g^2}{16\pi^3} (1-x) \!\!\int d^2{\vec{k}}_\perp  \biggl[\frac{(m+xM)^2}{[({\vec{k}}^{\prime 2}_\perp+u(\lambda^2))({\vec{k}}^2_\perp+u(\lambda^2)]} +\frac{k^{\prime 1}-ik^{\prime 2}}{({\vec{k}}^{\prime 2}_\perp+u(\lambda^2))}\cdot\frac{k^{ 1}+ik^{ 2}}{({\vec{k}}^{ 2}_\perp+u(\lambda^2))}\biggr]\label{eq:HGPDs1} \ee
and
\be E(x,0,-{\vec \Delta}^2_\perp)&=&\frac{-2M}{\Delta^1-i\Delta^2}\int \frac{d^2{\vec{k}}_\perp}{16\pi^3}\times
\biggr[\psi^{\uparrow\,*}_{+\frac{1}{2}}(x,{\vec{k}}^\prime_\perp)\psi^\downarrow_{+\frac{1}{2}}(x,{\vec{k}}_\perp)+\psi^{\uparrow\,*}_{-\frac{1}{2}}(x,{\vec{k}}^\prime_\perp)\psi^\downarrow_{-\frac{1}{2}}(x,{\vec{k}}_\perp)\biggr]\label{eq:EGPDs} \\
&=&\frac{-2Mg^2}{16\pi^3(\Delta^1-i\Delta^2)} \!\int  d^2{\vec{k}}_\perp  \times \frac{(Mx+m)(1-x)}{[({\vec{k}}^{\prime 2}_\perp+u(\lambda^2))({\vec{k}}^2_\perp+u(\lambda^2)]}\biggr[  (k^1-ik^2)-(k^{\prime 1}-ik^{\prime 2})\biggr],\label{eq:EGPDs1} \ee
where ${\vec{k}}^\prime_\perp ={\vec{k}}_\perp + (1-x)\vec{\Delta}_\perp$ is the relative transverse momentum of the quark in the final state of a nucleon and $u(\lambda^2)$ is defined in Eq.~(\ref{eq:phiu}). Since some of the above ${\vec k}_\perp$-integrals diverge, a manifestly Lorentz-invariant Pauli-Villars regularization is applied to regularize the divergent pieces of the integrals. For this purpose, $\lambda^2 \rightarrow \Lambda^2$$\,(=10 \lambda^2)$ is employed throughout the work presented in this paper.

Using Eq.~(\ref{eq:Jirelation}) and the GPDs available in Eqs.~(\ref{eq:HGPDs1}) and (\ref{eq:EGPDs1}), one can calculate
the quark OAM for a nucleon polarized in the $+z$ direction as
\be \begin{split}  L^z &= \frac{1}{2}\!\int_0^1\!\!dx\, \left[xH(x,0,0)+xE(x,0,0)-\Delta q (x) \right],\\ \text{where}\\ \Delta q(x)&=\int \!\frac{d^2{\vec k}_\perp}{16\pi^3}
\left[ \left|\psi_{+\frac{1}{2}}^\uparrow(x,{\vec k}_\perp) \right|^2 -
\left|\psi_{-\frac{1}{2}}^\uparrow(x,{\vec k}_\perp)\right|^2\right], \label{eq:LSDQM} \end{split}
\ee
and the quark spin angular momentum reads \cite{Brodsky:1980zm, Brodsky:2000ii}
\be  S = \frac{1}{2} \int_0^1 \! \Delta q(x) dx = \frac{g^2}{32 \pi^2} \int_0^1 (1-x) \times \left [ (Mx+m)^2 \left (\frac{1}{u(\lambda^2)} -\frac{1}{u(\Lambda^2)}\right )- \ln \left ( \frac{u(\Lambda^2)}{u(\lambda^2)}\right ) \right ] dx.
\label{eq:S} \ee

Now, from Eqs.~(\ref{eq:Jqt}) and (\ref{eq:TAMnaive}), one can define GPDs and TAM in impact parameter space $b_\perp$ as \cite{Burkardt:2003je}
\be  \widetilde{J}({\vec b}_\perp)= \frac{1}{2} \int_0^1 x [\mathcal{H}(x,{\vec b}_\perp) + \mathcal{E}(x,{\vec b}_\perp)]dx, \label{eq:Jb}\ee
where\be \begin{split}   \mathcal{H}(x,{\vec b}_\perp)&=
\int \frac{d^2{\vec \Delta}_\perp}{(2\pi)^2}
e^{-i {\vec \Delta}_\perp \cdot {\vec b}_\perp }
H(x,0,-\vec{\Delta}_\perp^2) \\
&=\biggr[\psi_{+\frac{1}{2}}^{\uparrow *}(x,{\vec r}_\perp) \psi_{+\frac{1}{2}}^{\uparrow}(x,{\vec r}_\perp) +\psi_{-\frac{1}{2}}^{\uparrow *}(x,{\vec r}_\perp) \psi_{-\frac{1}{2}}^{\uparrow}(x,{\vec r}_\perp) \biggr] \times\frac{1}{(1-x)^2} \label{eq:Hb} \end{split} \ee
and
\be \begin{split} \mathcal{E}(x,{\vec b}_\perp) &=
\int \frac{d^2{\vec \Delta}_\perp}{(2\pi)^2}
e^{-i {\vec \Delta}_\perp \cdot {\vec b}_\perp }
E(x,0,-\vec{\Delta}_\perp^2)\\
\frac{(-i\frac{\partial}{\partial b^x}-\frac{\partial}{\partial b^y})}{2M} \mathcal{E}(x,{\vec b}_\perp)&= \biggr[\psi_{+\frac{1}{2}}^{\uparrow *}(x,{\vec r}_\perp) \psi_{+\frac{1}{2}}^{\downarrow}(x,{\vec r}_\perp)+\psi_{-\frac{1}{2}}^{\uparrow *}(x,{\vec r}_\perp) \psi_{-\frac{1}{2}}^{\downarrow}(x,{\vec r}_\perp) \biggr]\frac{1}{(1-x)^2}  \label{eq:Eb}
 ,\end{split}\ee
 where ${\vec b}_\perp$ and ${\vec r}_\perp$ are related by ${\vec b}_\perp = (1-x){\vec r}_\perp$ \cite{Burkardt:2003je}. This relation is recently discussed in Ref.~\cite{Mondal:2016xsm} using LFWFs constructed from the soft-wall ADS/QCD prediction.

Here, in order to describe distributions in impact parameter space, we introduce LFWFs in the impact parameter space $b_\perp$ as
\be \begin{split}
\psi^{\uparrow\downarrow}_{\pm\frac{1}{2}}(x,{\vec b}_\perp) &\equiv \frac{1}{2\pi(1-x)}
\int d^2{\vec k}_\perp e^{-i\frac{ {\vec k}_\perp \cdot
{\vec b}_\perp}{1-x}} \psi^{\uparrow\downarrow}_{\pm\frac{1}{2}}(x,{\vec k}_\perp),\\  \psi^{\uparrow\downarrow}_{\pm\frac{1}{2}}(x,{\vec r}_\perp)&\equiv \frac{1}{2\pi}\!
\int d^2{\vec k}_\perp e^{-i {\vec k}_\perp \cdot
{\vec r}_\perp} \psi^{\uparrow\downarrow}_{\pm\frac{1}{2}}(x,{\vec k}_\perp).
  \label{eq:Ftrelation}
\end{split}
\ee
The factor $\frac{1}{1-x}$ in the exponent
accounts for the fact that the variable ${\vec k}_\perp$
is Fourier-conjugate to ${\vec r}_\perp ={\vec r}_{\perp 1}-{\vec r}_{\perp 2}$, the displacement  between the active quark
and the spectator (scalar). The prefactor  $\frac{1}{(1-x)}$ also ensures the proper normalization of the wave functions.
\be \begin{split}  \int|\psi^{\uparrow\downarrow}_{\pm\frac{1}{2}}(x,\vec{b}_\perp)|^2 \,d^2 \vec{b}_\perp\,& = \int|\psi^{\uparrow\downarrow}_{\pm\frac{1}{2}}(x,\vec{r}_\perp)|^2 \,d^2 \vec{r}_\perp =\int|\psi^{\uparrow\downarrow}_{\pm\frac{1}{2}}(x,\vec{k}_\perp)|^2 \,d^2\vec{k}_\perp. \end{split} \ee

Inserting Eqs.~(\ref{eq:Hb}), (\ref{eq:Eb}), (\ref{eq:Ftrelation}), and (\ref{eq:SDQM}) into  Eq.~(\ref{eq:Jb}) yields
\be \begin{split} \widetilde{J}(b_\perp)= \frac{g^2}{32 \pi^3}\int_0^1 \frac{x}{1-x} \times \biggr [(Mx+m)^2 \big [\bes_0(Z)\big ]^2  +u(\lambda^2) \big [\bes_1(Z)\big ]^2  \biggr ]dx +\frac{2 Mg^2}{32 \pi^3}\int_0^1  x (Mx+m)\big [\bes_0(Z)\big ]^2 dx ,\label{eq:TAMnaive1}
\end{split}\ee
where $Z= \left |\frac{b_\perp}{1-x}\right |\sqrt{u(\lambda^2)}$, $u(\lambda^2)$ is defined in Eq.~(\ref{eq:phiu}), and $\bes_n(Z)$ is the modified Bessel function of second kind.

Using Eqs.~(\ref{eq:Jqt}), (\ref{eq:TAMnaive}), (\ref{eq:HGPDs1}), (\ref{eq:EGPDs1}), and (\ref{eq:Jb}), it is straight forward to verify the following relation numerically for a consistency check of our expressions.
\be  \int d^2b_\perp  \widetilde{J}(b_\perp) = \frac{1}{2}\int dx\,x\left[ H(x, 0, 0) +E(x, 0, 0)\right].
\ee

Using the LFWFs available in Eq.~(\ref{eq:Ftrelation}), one can evaluate the spin angular momentum distribution in the transverse plane $(b_\perp)$ for a nucleon polarized in the $+z$ direction as
\be \hspace*{-.3cm}
S(b_\perp) = \frac{1}{2}\int_0^1 \!\! dx \left[ \left|
\psi^\uparrow_{+\frac{1}{2}}(x,{\vec b}_\perp)\right|^2
- \left|\psi^\uparrow_{-\frac{1}{2}}(x,{\vec b}_\perp)\right|^2\right], \label{eq:spinb}
\ee

where \be \begin{split} \left|\psi^\uparrow_{+\frac{1}{2}}(x,\vec{b}_\perp)\right|^2 = \frac{g^2}{16 \pi^3} \frac{(Mx+m)^2}{(1-x)} \big [\bes_0(Z)\big ]^2  , \quad \label{eq:pluspsi}   \left|\psi^\uparrow_{-\frac{1}{2}}(x,\vec{b}_\perp)\right|^2=\frac{g^2}{16\pi^3} \frac{u(\lambda^2)}{(1-x)}\left [\bes_1(Z)\right ]^2. \label{eq:minuspsi} \end{split}\ee

If one assumes that $\widetilde{J}(b_\perp )$, $\rho_J^{\text{PG}}(b_\perp )$, and $\rho_J^{\text{IMF}}(b_\perp )$  can be interpreted as TAM densities then the differences
\be
L^{\text{naive}}(b_\perp)\equiv& \widetilde{J}(b_\perp)-
S(b_\perp) \label{eq:naiveL}, \\
 L^{\text{PG}}(b_\perp)\equiv& \rho_J^{\text{PG}}(b_\perp)-
S(b_\perp) \label{eq:rhoGoekeL}, \quad \text{and} \\
L^{\text{IMF}}(b_\perp)\equiv& \rho_J^{\text{IMF}}(b_\perp)-
S(b_\perp) \label{eq:rhoIMFL}
\ee
would have to represent the respective OAM densities.

In the following section, we will investigate if that is indeed the case by comparing these distributions with the one calculated directly from LFWFs in impact parameter space using the JM definition.
\begin{figure*}
\includegraphics[scale=0.73]{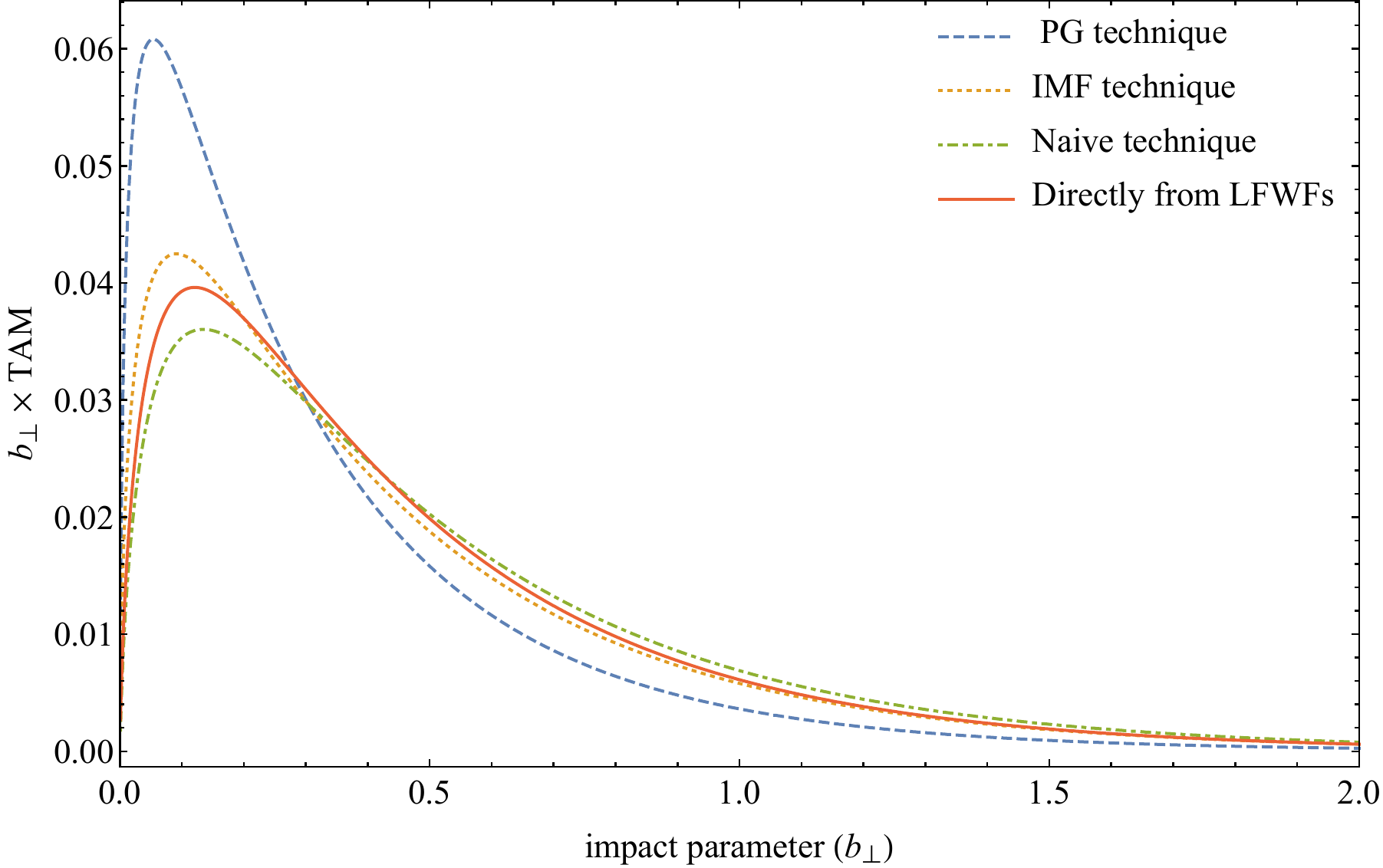}
\caption{Total angular momentum (TAM) distribution of the quark in the scalar diquark model for a nucleon polarized in the $+z$ direction. Solid line: ${\cal L}(b_\perp)+S(b_\perp)$ , directly from LFWFs [Eq.~(\ref{eq:LCoam}) +Eq.~(\ref{eq:spinb})] , Dashed line: $\rho_J^{\text{PG}}(b_\perp)$, Polyakov-Goeke (PG) technique [Eq.~(\ref{eq:TAMGoeke})] , Dotted line: $\rho_J^{\text{IMF}}(b_\perp)$, IMF technique [Eq.~(\ref{eq:TAMIMF})], Dash-dotted line: ${\widetilde J}(b_\perp)$, naive technique [Eq.~(\ref{eq:TAMnaive1})]. The plots are in units of $\frac{g^2}{16 \pi}$ } \label{fig:J}
\end{figure*}
\begin{figure*}
\includegraphics[scale=0.74]{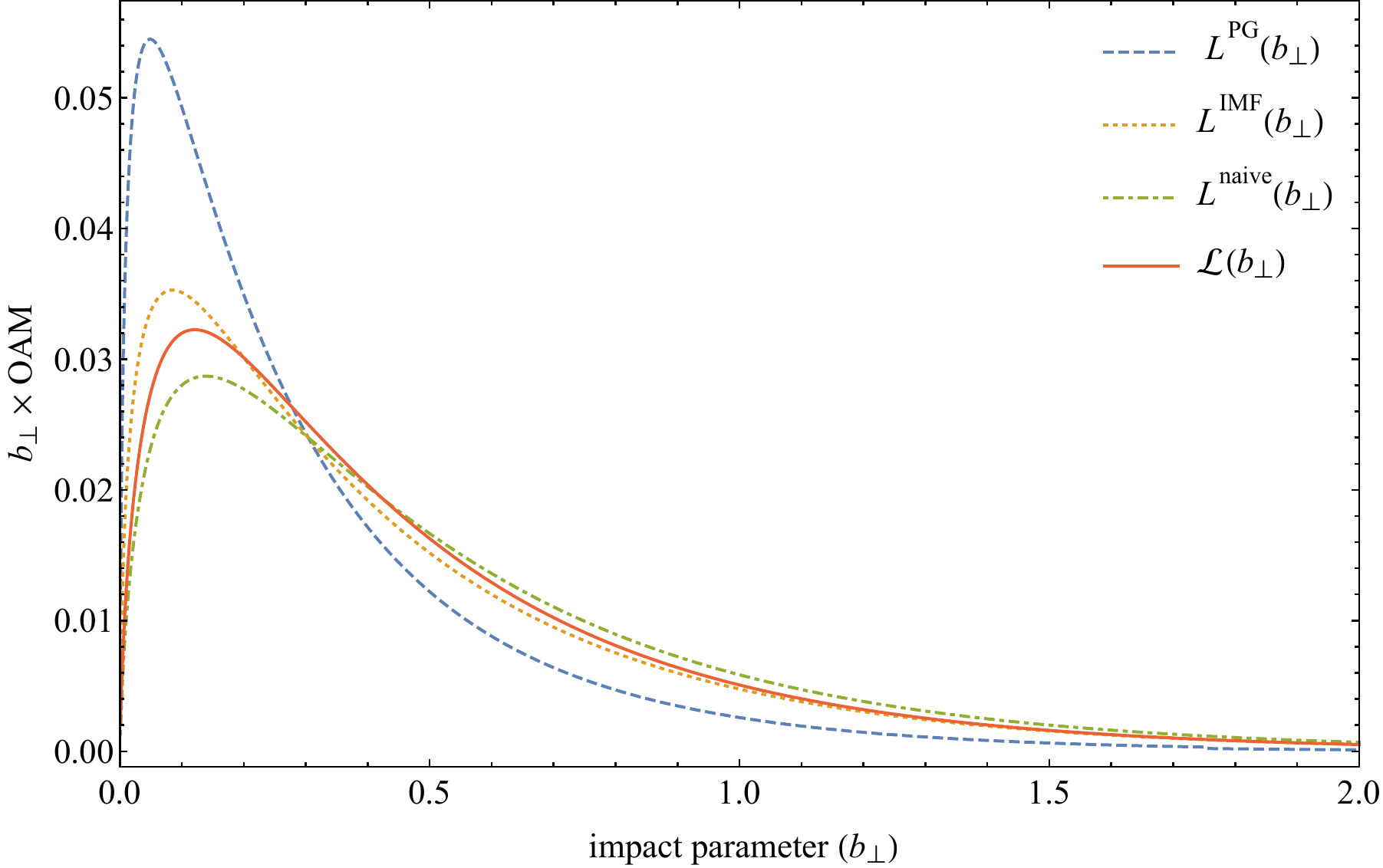}
\caption{Orbital angular momentum (OAM) distribution of the quark in the scalar diquark model for a nucleon polarized in the $+z$ direction. Solid line: ${\cal L}(b_\perp)$ , directly from LFWFs [Eq.~(\ref{eq:LCoam})], Dashed line: $L^{\text{PG}}(b_\perp)$, Polyakov-Goeke (PG) technique [Eq.~(\ref{eq:rhoGoekeL})] , Dotted line: $L^{\text{IMF}}(b_\perp)$, IMF technique [Eq.~(\ref{eq:rhoIMFL})], Dash-dotted line: $L^{\text{naive}}(b_\perp)$, naive technique [Eq.~(\ref{eq:naiveL})]. The plots are in units of $\frac{g^2}{16 \pi}$ } \label{fig:L}
\end{figure*}
\subsection{Impact Parameter Space Distribution Directly from Light-Front Wavefunctions:}
With the LFWFs available in Eq.~(\ref{eq:SDQM}), one can compute the orbital angular momentum ${\cal L}^z$ of the `quark' for a `nucleon' polarized in the $+z$ direction directly
as \cite{Harindranath:1998ve, Burkardt:2008ua, Bashinsky:1998if,Jaffe:1989jz}
\be \begin{split}
{\cal L}^z = \int_0^1 dx \int \frac{d^2{\vec k}_\perp}{16\pi^3}
(1-x) \left|\psi_{-\frac{1}{2}}^\uparrow(x, {\vec k}_\perp)\right|^2 = \frac{g^2}{16 \pi^2} \int_0^1 (1-x)^2 \ln\left [\frac{u(\Lambda^2)}{u(\lambda^2)}\right ] dx.\label{eq:JaffeL}  \end{split}\ee
Manifestly Lorentz-invariant Pauli-Villars regularization is applied to evaluate the above integral.
It is straightforward to show
\be
{\cal L}^z = L^z \label{eq:BCresult}
\ee
as was expected since $L^z$ in the SDQM does not contain a vector potential, and therefore no gauge
related issues arise \cite{Burkardt:2008ua,Ji:2015sio}.
Likewise, one can define the OAM
density directly from the LFWFs available in
Eq.~(\ref{eq:Ftrelation}) as
\be \begin{split}
{\cal L}(b_\perp) =\int_0^1 dx\, (1-x)\left| \psi_{-\frac{1}{2}}^\uparrow
(x,{\vec b}_\perp)\right|^2 = \frac{g^2}{16\pi^3}\int_0^1 dx \,u(\lambda^2)\left [\bes_1(Z)\right  ]^2.
\label{eq:LCoam}
\end{split}\ee
Here,  ${\cal L}(b_\perp)$ represents the orbital angular momentum density for the active quark as a function of the distance from the
transverse center of momentum in a `nucleon' that is polarized in the $+z$ direction.

The TAM distribution of the active quark in the transverse plane is shown in Fig.~\ref{fig:J}. For the different techniques, obviously, the area under the graphs is the only feature that all four  distributions have in common, i.e.
\be \begin{split}
\int_0^\infty \!\!\!db_\perp b_\perp \widetilde{J}(b_\perp)\! =\!\!\int_0^\infty \!\!\! db_\perp b_\perp  \rho^{\text{PG}}_J(b_\perp) =\!\int_0^\infty  \!\!\! db_\perp  b_\perp \rho^{\text{IMF}}_J(b_\perp)=\int_0^\infty \!\!\! db_\perp b_\perp (\mathcal{L}(b_\perp) +S(b_\perp)) .\label{eq:Jcheck}  \end{split}\ee

Similarly, the OAM distributions of the active quark in the transverse plane are shown in Fig.~\ref{fig:L}; also in this case, the area under the graphs for all four  distributions is the only common feature, i.e,
\be \begin{split} \int_0^\infty \!\!\!db_\perp b_\perp {L}^{\text{naive}}(b_\perp)\! =\!\!\int_0^\infty \!\!\! db_\perp b_\perp  {L}^{\text{PG}}(b_\perp) =\!\int_0^\infty  \!\!\! db_\perp  b_\perp {L}^{\text{IMF}}(b_\perp)=\int_0^\infty \!\!\! db_\perp b_\perp \mathcal{L}(b_\perp)=\frac{{\cal L}^z}{2\pi}.  \label{eq:Lcheck}  \end{split}\ee
Eqs.~(\ref{eq:Jcheck}) and (\ref{eq:Lcheck})  ensure that the TAM/OAM does not change regardless of whether one uses a different frame or technique to perform the calculations.

The $b_\perp$- distributions of OAM presented in Fig.~\ref{fig:L} for the three different techniques [naive, Polyakov-Goeke (PG), and IMF] do not agree with the one associated with the definition introduced by Jaffe-Manohar. Therefore, these results clearly demonstrate that $L^{\text{naive}}(b_\perp)$, $L^{\text{PG}}(b_\perp)$, and $L^{\text{IMF}}(b_\perp)$ do not represent the orbital angular momentum distribution for a longitudinally polarized nucleon since $\mathcal{L}(b_\perp)$ already has that interpretation in momentum space. Furthermore, using the naive technique in the SDQM, we also conclude that the FT of $J(t)$ does not represent the distribution of angular momentum in the transverse plane regardless of whether the FT is 2-dimensional or 3-dimensional. All three techniques discussed above are associated with the FT of  $J(t)$. While $J(t)$ itself indeed is identified with the $2^{\text{nd}}$ moment of GPDs in the limiting case $t\rightarrow 0$, our investigation exhibits three different possibilities of relating the $t$-dependence of GPDs to the angular momentum distributions in the transverse plane. None of them turns out to yield the distribution one would expect from the Jaffe-Manohar definition for a longitudinally polarized nucleon.
\section{Discussion}
\subsection{Naive technique}
It was demonstrated using the SDQM  that, although $J(t)$ yields the $z$-component of the total angular momentum of the quarks for a nucleon polarized in the $+z$ direction in the limit $t\rightarrow 0$, the 2-dimensional FT of its $t$-dependence does not yield the distribution of angular momentum in the transverse plane.

This result is best understood by recalling that Lorentz/rotational-invariance was heavily used
in Ref.~\cite{Ji:PRL}
for deriving Eq.~(\ref{eq:Jirelation}) as it restricts the allowed tensor structure.
In Ref. \cite{Burkardt:2005hp},
Ji's angular momentum sum rule was rederived by considering the transverse deformation
of parton distributions in a transversely polarized nucleon, and in several steps of the derivation
rotational invariance was used for rotations that mix `longitudinal' and `transverse' directions.
When one considers distributions in the transverse plane, rotational invariance is no longer fully applicable. This is analogous to the observation that the unintegrated Ji relation , i.e. $J(x)\equiv  \frac{x}{2} [ H(x,0, 0)+E(x,0, 0)]$  is not the $x$- distribution of  $J^z(x)$ for a longitudinally polarized nucleon \cite{Burkardt:2008ua}.
\subsection{Polyakov-Goeke (PG) technique  and IMF technique}
In hadron spin structure studies, the total angular momentum of a quark is decomposed into spin and orbital parts, and the spin distribution of the quark in the transverse plane can be obtained using a 2-dimensional FT of axial form factors. To study angular momentum distribution in the transverse plane, one may be tempted to interpret the two distributions (densities), proposed in Eq.~(\ref{eq:TAMGoeke}) and Eq.~(\ref{eq:TAMIMF}), as a sum of spin and orbital angular momentum distributions (densities).
In particular the observation that $\rho_J^{\text{IMF}}(b_\perp)$ differs from the light-front wavefunctions
based result may thus appear surprising.
However, both proposed densities have in common that they are based on the symmetric energy momentum tensor $T^{\mu \nu}$. On the other hand, the symmetric energy momentum tensor $T^{\mu \nu}$ can be expressed in terms of canonical energy momentum tensor $\mathcal{T}^{\mu \nu}$ and spin current $S^{\mu}_{\nu \lambda}$\cite{Belinfante, Rosenfeld}. The total angular momentum density, which preserves the interpretation as a sum of spin and orbital angular momentum densities, is based on the canonical energy momentum tensor. Therefore, it is important to illustrate the role of the total divergence term  available in $T^{\mu \nu}$.

 The symmetric energy momentum tensor for the massless Dirac particle can be expressed as
\be T^{\mu \nu} = \frac{i}{2}\bar{q}\left(
\gamma^\mu \stackrel{\leftrightarrow}{D^\nu}+
\gamma^\mu \stackrel{\leftrightarrow}{D^\nu} \right)q. \ee
Using the equations of motion (valid in matrix elements), one finds
\be \begin{split}
\bar{q}\gamma^x\stackrel{\leftrightarrow}{D^+}q =\bar{q}\gamma^+ \stackrel{\leftrightarrow}{D^x} q
 - \partial^y\left(
\bar{q}\gamma^+ \gamma^x\gamma^y q\right)
+\frac{1}{4}\partial_-\left[q_-^\dagger \gamma^0 \gamma^x\gamma^+\gamma^-q_+ -
q_+^\dagger \gamma^0\gamma^x\gamma^-\gamma^+q_-\right].
\end{split}\ee
Inserting into the total angular momentum density, one finds
\be \begin{split}
xT^{+y}-yT^{+x}&= x\bar{q}\gamma^+ i\!\!\stackrel{\leftrightarrow}{D^y}\! q-
y\bar{q}\gamma^+i\!\!\stackrel{\leftrightarrow}{D^x}\! q
+\bar{q}\gamma^+ i\gamma^x\gamma^y q +i\partial_x\!\left(x \bar{q}\gamma^+ \gamma^x\gamma^y q\right)
+i\partial_y\!\left(y \bar{q}\gamma^+ \gamma^x\gamma^y q\right) \\
&+\frac{i}{4}\partial_-\left[x\bar{q}\gamma^y\gamma^+\gamma^-q -
y\bar{q} \gamma^x\gamma^+\gamma^-q\right], \label{eq:totalderiv}
\end{split}\ee
therefore, there are two terms that together have the physical interpretation as an orbital angular
momentum density, one term that represents the spin density and a total derivative term. While the presence of these total derivative terms has no consequences for the integrated quantities, they cause a profound dilemma when attempting to study angular momentum densities: Though $xT^{+y}-yT^{+x}$ seems to be a perfect candidate for the total angular momentum density, one has to be careful not to interpret that density as a simple sum of orbital angular momentum density and spin density. This statement may sound paradoxical but is due to the presence of terms that are total derivatives and thus do not contribute to the overall angular momentum. Nevertheless, these contributions play an important role while attempting to study the angular momentum density in the transverse plane. Note that, although the issue of total derivative terms was raised and discussed in Ref.~\cite{Leader:2013jra}, it was illustrated explicitly in this paper using model calculations.
\section{Acknowledgements}
This work was supported by the US Department of Energy under Grant No. DE-FG03-95ER40965. LA was supported in part by the US Department of Energy under Grant No. DE-FG02-87ER40371. MB would like to thank C. Lorc\'e, B. Pasquini, and M. Polyakov for useful discussions.

\end{document}